# Vers des supercondensateurs plus performants: quand expériences et simulations permettent d'élucider les mécanismes à l'échelle nanométrique

**Résumé**

Les supercondensateurs sont des dispositifs de stockage de l'énergie électrique permettant notamment de délivrer une forte puissance. Ils sont constitués d'électrodes de carbone poreux plongées dans un électrolyte concentré. La charge est stockée par adsorption des ions sur la surface des électrodes. Les carbones nanoporeux permettent de stocker une plus grande quantité d'électricité grâce à un mécanisme de charge différent des carbones possédant des pores de plus grande taille. Ce mécanisme de charge a été récemment caractérisé en combinant des expériences *in situ* (Résonance Magnétique Nucléaire, microbalance à quartz) et des simulations moléculaires.

# Towards more efficient supercapacitors: When experiments and simulations uncover the mechanisms on the nanometer scale

**Abstract**

Supercapacitors are energy storage devices able to deliver electricity with a high power. They consist of porous carbon electrodes in a concentrated electrolyte. Charged is stored by the adsorption of ions at the electrode surface. Nanoporous carbons allow to store more electricity thanks to a charging mechanism that differs from carbons with larger pores. This charge storage mechanisms has recently been uncovered by combining *in situ* experiments (Nuclear Magnetic Resonance, Electrochemical Quartz Microblance) and molecular simulations.

**Auteurs**

Benjamin Rotenberg[1,2], Mathieu Salanne[1,2], Patrice Simon[2,3]

1 Sorbonne Universités, UPMC Univ Paris 06, CNRS, Laboratoire PHENIX, 4 place Jussieu, 75005 Paris

2 RS2E (Réseau sur le Stockage Electrochimique de l'Energie), FR CNRS 3459, 80039 Amiens Cedex, France

3 CIRIMAT, Université de Toulouse, CNRS, INPT, UPS, 118 route de Narbonne, 31062 Toulouse Cedex 9, France

Benjamin Rotenberg, chargé de recherche CNRS, et Mathieu Salanne, maître de conférences, au laboratoire Physicochimie des Electrolytes et Nanosystèmes Interfaciaux (PHENIX, UMR 8234) à l'UPMC (Université Pierre et Marie Curie, Paris). Patrice Simon, professeur au Centre Inter-universitaire de Recherche et d'Ingénierie des Matériaux  CIRIMAT, à l'Université Paul Sabatier (Toulouse).

# I. Introduction

Le réchauffement de la planète, les réserves limitées en combustibles fossiles et la pollution des villes (les transports sont responsables de 30 % des émissions de $CO_2$) montrent, entre autres, combien il est important de se tourner vers une utilisation intensive et efficace des énergies renouvelables (ENR) et de trouver des solutions innovantes pour faciliter le passage progressif du véhicule thermique aux véhicules électriques. L'intermittence des énergies renouvelables ainsi que la nécessité d'embarquer une quantité suffisante d'électricité dans les véhicules électriques afin d'assurer une grande autonomie, font du développement de nouvelles technologies pour le stockage de l'énergie un des défis majeurs des vingt prochaines années: il s'agit d'une étape indispensable pour mieux gérer les ressources en énergie de notre planète. S'il existe une grande variété de solutions de stockage à grande échelle (stockage hydraulique, air comprimé, volants d'inertie…), la technologie la plus largement utilisée reste incontestablement le stockage électrochimique avec les batteries et les supercondensateurs [1]. Ce succès s'explique par l'avantage considérable qu'ils apportent par rapport à d'autres solutions : la mobilité. La mise au point de générateurs électrochimiques performants revêt donc une importance toute particulière [2,3].

Comme le montre la **Figure 1a**, les batteries permettent de stocker ou délivrer de grandes quantités d'énergie (200 Wh/kg) pendant des temps longs, typiquement de plusieurs heures ou dizaines d'heure. Cependant, de par le mécanisme de stockage des charges impliquant la rupture / création de liaisons chimiques par des réactions d'oxydo-réduction, elles restent limitées en terme de puissance. A l'inverse, les condensateurs diélectriques classiques ne stockent que peu d'énergie mais sont capables de la restituer en quelques millisecondes, permettant ainsi délivrer de grandes puissances.

Les supercondensateurs sont des systèmes intermédiaires entre les condensateurs classiques et les batteries. Ils peuvent fournir des densités de puissance très élevées (> 10 kW/kg) avec une densité d'énergie modeste (6 Wh/kg), ce qui définit une constante de temps allant de quelques secondes à quelques dizaines de secondes. Ces performances sont liées au mode de stockage de la charge : il n'y a en effet aucune réaction redox dans les supercondensateurs car le stockage est réalisé de façon électrostatique, par adsorption des ions d'un électrolyte à la surface de carbones poreux de surface très développée. Ce mécanisme peut être décrit en première approche en utilisant la notion de capacité de double couche électrochimique (**Figure 1b**), suivant le modèle proposé par Helmholtz en 1853 :

$$C = \frac{\varepsilon_0 \varepsilon_r A}{d} \qquad (1)$$

où $\varepsilon_0$ la permittivité du vide, $\varepsilon_r$ est la constant diélectrique de l'électrolyte, $d$ l'épaisseur de la double couche électrochimique (distance de séparation des charges), et $A$ l'aire de l'électrode. On notera qu'excepté $\varepsilon_0$, ces grandeurs sont difficiles à définir à l'échelle locale des interfaces. La capacité de cette double couche électrochimique est de l'ordre de 10 à 20 µF/cm². En remplaçant une électrode plane par un matériau poreux de grande surface spécifique comme le carbone activé (1000 à 2000 m²/g, **Figure 1c**), on atteint des capacités de plus de 100 F/g de carbone. La tension maximum de ces cellules est limitée par la décomposition de l'électrolyte : 2,7 à 3 V en électrolyte non-aqueux. Le stockage des charges en surface explique la grande puissance de ces systèmes par rapport aux batteries. L'absence de variation de volume dans les électrodes au cours des cycles de charge/décharge (la charge restant en surface) permet aux supercondensateurs d'atteindre des cyclabilités de plusieurs millions de cycles à température ambiante, soit bien plus que pour les batteries (typiquement quelques centaines). Enfin, l'utilisation de solvants comme l'acétonitrile permet un fonctionnement entre -40°C et +70°C. Rappelons toutefois que la densité d'énergie est environ 30 fois plus faible que celle des batteries.

Les supercondensateurs sont utilisés pour deux applications principales, qui sont la fourniture de pics de puissance et la récupération de l'énergie ; pour cette dernière, c'est la vitesse de recharge qui est exploitée [2]. On les retrouve en petit format (cellule de capacité de moins de 100F) en électronique de puissance. Les formats plus importants (capacité de plus de 100F par cellule) sont utilisés par exemple dans l'aéronautique (A380), l'automobile, les tramways et les bus (fonction stop and start et récupération de l'énergie de freinage), ou encore les grues portuaires (récupération de l'énergie potentielle)… Des applications récentes utilisent les supercondensateurs pour la traction électrique dans les bus qui font des arrêts réguliers. L'autonomie limitée (quelques minutes) reste suffisante pour rouler en mode tout électrique entre deux arrêts, et la recharge se fait en moins d'une minute pendant l'échange de passagers. Ils permettent également, en association avec les batteries, d'augmenter la durée de vie de ces dernières en fournissant les pics de puissance qui sont les plus contraignants pour la batterie. Jusqu'en 2005, le modèle classique utilisé pour décrire l'adsorption des ions dans les pores des carbones prévoyait que seuls les pores de taille comprise entre 2 et 10 nm (les mésopores) permettaient un stockage efficace des ions ; la plupart des travaux étaient donc orientés vers la synthèse de carbones mésoporeux pour maximiser la capacitance. La découverte de l'augmentation de la capacité dans les nanopores (de taille inférieure à 1 nm, c'est-à-dire inférieure à la taille des ions solvatés) a conduit à complètement repenser l'adsorption des ions dans les pores confinés, et donc la charge de la double couche à l'échelle nanométrique [4]. Du point de vue pratique, la première conséquence a été l'utilisation, dans les systèmes commerciaux, de carbones microporeux dont tout le volume poreux provient de pores de taille inférieure à 2 nm [5]. Du point de vue scientifique, il a fallu développer de nouvelles techniques, expérimentales et théoriques, pour essayer de comprendre l'organisation des ions de l'électrolyte dans les pores nanométriques et sub-nanométriques des carbones pour essayer d'expliquer ces capacités élevées dans ces pores confinés [6]. Ces 5 à 10 dernières années, les techniques de caractérisation *in situ* par diffusion aux petits angles des rayons X (SAXS) et des neutrons (SANS), ainsi que les approches théoriques par dynamique moléculaire classique ou *ab initio* [7] ont permis de faire des avancées importantes dans le domaine. Le développement de techniques électrochimiques avancées, comme la microbalance à quartz électrochimique couplée à des techniques spectroscopiques comme la RMN, ont également été à l'origine de progrès notables, en contribuant à comprendre le transport et l'adsorption des ions dans les pores. Nous développons ici le sujet abordé succinctement dans un récent numéro de l'Actualité Chimique consacré à la transition énergétique [8].

## II. Modèles structuraux des carbones nanoporeux

Les carbones nanoporeux sont généralement constitués d'unités graphitiques nanométriques présentant des défauts, avec des pores de tailles couvrant les micropores (taille < 2 nm) aux mesopores (taille entre 2 et 50 nm). Contrairement aux pores en fentes ou aux nanotubes, ils ne présentent aucun ordre tri-dimensionnel à longue portée [Figure 2]. Puisque les performances électrochimiques des carbones sont déterminées par l'interface entre le carbone poreux et l'électrolyte, il est nécessaire de pouvoir caractériser de façon fiable et précise le réseau poreux. Les surfaces spécifiques (surface par unité de volume ou de masse de matériau) ou les distributions de tailles de pores sont généralement obtenues à partir de mesures d'adsorption de gaz, via des modèles théoriques. Cependant ces méthodes présentent des limitations importantes lorsqu'on chercher à les appliquer aux carbones microporeux.

Le choix de la sonde moléculaire est essentiel, car certains pores peuvent être inaccessibles aux molécules de gaz et l'être quand même pour les ions de l'électrolyte, selon les conditions thermodynamiques. Ceci est particulièrement vrai pour l'adsorption de $N_2$ à 77 K, qui ne donne pas des mesures précises pour les ultra-micropores (< 0,7 nm) dans les carbones. L'utilisation de l'argon, plus petit et sans quadrupôle, et ce à plus haute température (87 K), combinée à une analyse théorique plus simple des données expérimentales, est plus adaptée dans ce cas. L'adsorption de fluides super- ou sous-critiques ($CO_2$) à température ambiante offre une approche complémentaire pour caractériser la surface accessible dans des conditions plus proches de celle d'utilisation.

Outre la sonde moléculaire et les conditions thermodynamiques, un autre défi doit être surmonté, car les propriétés microscopiques doivent être déduites des isothermes d'adsorption à travers un modèle approprié. L'analyse BET (Brunauer-Emmett-Teller) n'est pas recommandée pour les carbones microporeux, comme l'a confirmé récemment l'IUPAC [9], car elle sous-estime les ultra-micropores et surestime les micropores de plus grande taille. Il est nécessaire d'avoir recours à des techniques plus élaborées telles que la méthode SPE (substracting pore effect) [10]. Si l'on souhaite accéder, au-delà de la surface spécifique, à une mesure des tailles de pores et de leur distribution, on peut avoir recours à une analyse de type théorie de la fonctionnelle de la densité (Quenched Solid Density Functional Theory) [11].

Tout ceci illustre la difficulté de mesure la surface des carbones nanoporeux. En réalité, le concept même de surface est à manier avec précaution, car il dépend de l'approche (notamment, la sonde choisie) pour la mesurer. Il est donc pour cela nécessaire de compléter ces techniques d'adsorption par une caractérisation directe de la structure telle que la diffusion de rayons X aux petits angles (SAXS) ou la résonance magnétique nucléaire (RMN), et avoir ainsi accès au rapport surface sur volume accessible aux ions, c'est-à-dire dans des conditions pertinentes pour les applications électrochimiques. Malgré toutes les limitations évoquées ci-dessus, il reste possible de discuter l'évolution de la capacité par unité de surface en fonction de la taille de pore à partir des mesures d'adsorption d'argon, pour comparer différents carbones. On préférera toutefois les capacités gravimétriques ($F.g^{-1}$ d'électrode) ou volumétriques ($F.cm^{-3}$ d'électrode) qui sont mesurables directement, sans avoir recours à des considérations théoriques ou structurales.

Il est beaucoup plus délicat de caractériser la "vraie" structure des carbones nanoporeux, car il n'est à ce jour pas possible de le faire à partir d'approches purement expérimentales. On a ainsi recours à des combinaisons modélisation/expérience. Par exemple, la diffraction des rayons X et le SAXS permettent d'obtenir des informations structurales à courte et longue distance. Mais l'on obtient en général des informations structurales à 1D, et le passage à la structure 3D se fait souvent en recourant à des simulations de Monte Carlo hybride inverse [12]. Une approche combinant RMN, rayons X, spectroscopie Raman et simulation sur réseau a ainsi récemment permis d'estimer la taille des domaines graphitiques dans les carbones poreux, tandis que des simulations de trempe de dynamique moléculaire ont permis d'obtenir des structures de carbone réalistes sans partir de données expérimentales [13].

Enfin, notons que la plupart des études ont pour l'instant porté sur la structure du carbone nanoporeux, alors que la capacité à prédire les paramètres structuraux cruciaux pour l'optimisation du stockage de charge nécessite également de comprendre la structure et la dynamique de l'électrolyte confiné.

**III. Mouillage des pores sans différence de potentiel**

Jusqu'à récemment, on croyait que la charge des supercondensateurs venait d'un mécanisme simple, à savoir l'entrée des ions dans le réseau de pores du carbone sous l'effet du champ électrique lors de l'application d'une différence de potentiel entre les électrodes. Les premières mesures de RMN *in situ* ont montré que ce n'est pas le cas. Même à faible concentration en électrolyte, on observe un décalage vers les basses fréquences du signal RMN, induit par les courants de cycle aromatique des domaines graphitiques et la susceptibilité magnétique du carbone [14]. Les résultats indiquent que les ions ainsi que le solvant sont bien présents dans les pores. La situation à 0 V est donc mieux décrite par l'interpénétration de deux structures hétérogènes: le carbone solide d'une part, et l'électrolyte d'autre part. Par des simulations de dynamique moléculaire, nous avons pu confirmer cette image, en montrant qu'un liquide ionique en contact avec une électrode de carbone nanoporeux entre spontanément dans les nanopores y compris en l'absence de différence de potentiel.

La spectroscopie RMN permet de plus de quantifier la concentration des ions adsorbés au sein de l'électrode [15]. Celle-ci est proportionnelle à la concentration dans le volume de l'électrolyte, ce qui confirme l'affinité des ions pour le carbone. Cependant, à l'échelle de temps de la mesure RMN, les ions diffusent dans la structure poreuse et sondent différents environnements, ce qui conduit à des spectres larges. Il n'est donc pas aisé de préciser les populations des différents sites d'adsorption par cette technique. On peut alors recourir à la diffusion au petits angles des rayons X ou de neutrons, en exploitant le contraste entre le carbone et l'électrolyte, pour préciser par exemple l'entrée ou non dans les pores de plus petite taille.

En combinant diffusion de neutrons et simulations moléculaires, Bañuelos *et al.* ont conclu qu'un liquide ionique à température ambiante (RTIL) couvrait la surface des pores d'un carbone à porosité hiérarchisé de manière uniforme, plutôt que de remplir certains pores complètement avant de passer à d'autres. Différentes observations ont été faites dans le cas des électrolytes aqueux, suggérant que la chimie de surface du carbone et la nature de l'électrolyte jouent un rôle important sur les propriétés de mouillage. Récemment, Kondrat et Kornyshev ont proposé d'utiliser des pores "ionophobes" pour la conception particulière d'une nouvelle génération de supercondensateurs [16]. Cette idée, dont la faisabilité expérimentale reste à démontrer, repose sur le fait que de tels pores se rempliraient seulement à haut potentiel, ce qui ouvre des perspectives intéressantes en termes de densité d'énergie et de vitesse de charge/décharge.

**IV. Désolvatation dans les nanopores**

La découverte que les ions d'un électrolyte pouvaient accéder et s'adsorber dans des pores de dimensions inférieures à la taille des ions solvatés a été le point de départ d'un grand nombre de travaux sur l'étude du confinement des ions dans les nanopores de carbone. Du point de vue expérimental tout d'abord, les résultats des caractérisations électrochimiques classiques (voltammétrie, chrono-potentiométrie ou encore spectroscopie d'impédance électrochimique) ont conduit à proposer que l'accès à ces nanopores se faisait en perdant une partie de leur cortège de solvatation. De plus, la capacité était fortement augmentée lorsque la taille des pores était du même ordre de grandeur que celle des ions nus [4,5]. Cependant, bien que ces résultats aient été obtenus avec des carbones à porosité contrôlée dans le domaine des micropores (<1,5 nm), ces techniques électrochimiques classiques ne donnent pas accès à des informations quantitatives sur, par exemple, le nombre de molécules de solvant perdues lors de l'adsorption dans les nanopores.

Le développement de techniques *in situ* comme la microbalance à quartz électrochimique (EQCM) s'est révélé être particulièrement important pour réaliser ce type de mesures [17]. La **Figure 3a** présente le schéma de principe d'une microbalance EQCM, dans laquelle un mélange de carbone poreux à étudier est placé sur un quartz piézoélectrique utilisé comme électrode de travail dans une cellule électrochimique. La variation de la fréquence de résonance du quartz est proportionnelle au changement de masse de l'électrode durant la polarisation. La **Figure 3b** montre la variation de masse (exprimée en nombre de mole d'anions et cations en divisant par la masse molaire des ions nus) en fonction de la charge lors de la polarisation d'une électrode de carbone. Les traits pointillés symbolisent la variation théorique de masse $\Delta m$ d'après la loi de Faraday :

$$\Delta m = \frac{Q}{F}\frac{M}{z} \qquad (2)$$

avec $Q$ la charge de l'électrode, $F$ la constante de Faraday, $M$ la masse molaire et $z$ la valence de l'espèce échangée. En première approche, on peut considérer que les contre-ions seuls s'adsorbent : les anions sont adsorbés pour des charges positives ($Q>0$), et les cations pour des charges négatives ($Q<0$). La **Figure 3b**, qui correspond à un électrolyte organique avec des carbones de taille de pores contrôlée de 1 nm, montre trois zones de pentes différentes, correspondants à des mécanismes différents. A faible charge, il y a un échange entre anions et cations : les contre-ions s'adsorbent tandis que les co-ions (charge de même signe que celle de l'électrode) sont expulsés. A charge plus importante, seuls les contre-ions s'adsorbent. De ces courbes, on peut déduire le nombre de molécules de solvant accompagnant les ions lors de leur adsorption dans les pores. En plus de mettre en évidence deux mécanismes de stockage des charges différents pour des polarisations positives et négatives (voir plus loin), la différence entre la variation de masse théorique et expérimentale de l'électrode pour les charges négatives a permis de calculer un nombre de solvatation de 3 pour les cations 1-éthyl-3-méthylimidazolium confinés dans les pores, alors que ce cation est normalement entouré de 8 molécules de solvant dans le même électrolyte non confiné [**15**]. Ces résultats sont la preuve expérimentale de la désolvatation partielle des ions dans les nanopores, et viennent confirmer les simulations par dynamique moléculaire qui ont également montré ce phénomène [**18**].

**V. Mécanisme de stockage de charge**

Si le principe de base du stockage de charge dans les supercondensateurs, à savoir l'adsorption d'ions à la surface de l'électrode, est bien établi, le mécanisme microscopique correspondant était bien moins clair jusqu'à récemment. La théorie de Gouy-Chapman-Stern (GCS) - qui prolonge le modèle de Helmholtz - reste la pierre angulaire de l'électrochimie théorique depuis plus d'un siècle ; elle prédit que près d'une surface étendue, la charge de l'électrode est compensée par la polarisation de l'électrolyte. La charge ionique et le potentiel électrostatique évoluent sur une distance caractéristique dite de Debye, de l'ordre de 1 à 10 nm en fonction de la concentration de l'électrolyte et la permittivité du solvant. Mais cette image est d'une utilité limitée dans le cas des supercondensateurs, à cause des effets de corrélation ionique à forte concentration en sel ou dans les liquides ioniques et des effets prononcés du confinement, qui est différent de la situation de l'électrode plane [19]. Les avancées récentes des techniques de simulations ont démontré la nécessité de décrire correctement la structure de l'électrolyte sur la surface.

Pour des surfaces complexes (électrodes poreuses), la simulation a mis en évidence un point important dans le cas du confinement extrême : lorsque la taille des pores est comparable à celle des ions, la charge de l'électrode est compensée par un déséquilibre entre les nombres de co- et contre-ions dans le pore. Dans un tel état "super-ionique", la violation de l'électro-neutralité dans le fluide interstitiel est compensée par l'apparition d'une charge opposée sur l'électrode, qui écrante la répulsion entre ions de même charge [20]. Plusieurs processus peuvent conduire à un excès global de contre-ions dans l'électrode: l'asdsorption de contre-ions, l'échange entre co- et contre-ions, ou la désorption de co-ions. Pour une combinaison donnée d'électrodes et d'électrolytes (nature des ions, présence et nature de solvant), l'un ou plusieurs de ces mécanismes peut être observé en fonction du potentiel. La simulation moléculaire, ainsi que les expériences de RMN, spectroscopie IR, SAXS et EQCM, suggèrent que l'échange d'ions est le processus le plus courant pour les faibles différences de potentiels, mais que pour des gros ions et/ou à fort potentiel l'adsorption des seuls contre-ions est aussi observée. La désorption des co-ions seule semble moins fréquente.

Des mesures récentes de RMN *in situ* avec un électrolyte dans un solvant ont montré que le mécanisme de stockage de charge peut varier en fonction de la polarisation : alors que pour les polarisations positives l'échange d'ion était observé, c'est l'adsorption des contre-ions qui dominait pour les polarisations négatives. Ainsi, plusieurs facteurs contribuent à l'excès de charge, tels que la taille et la mobilité relatives des co- et contre-ions, ou encore la réorganisation des ions sur plusieurs cycles charges/décharges. Comme nous l'avons déjà mentionné, tant l'adsorption de contre-ions que l'échange d'ions sont accompagnés d'entrées et sorties de molécules de solvant. Dans les liquides ioniques purs, les simulations moléculaires indiquent que l'échange d'ions a lieu sans changer le volume de liquide dans l'électrode (voir **Figure 4**) [**21**]. Cette conclusion reste à renforcer pour d'autres combinaisons de cations et d'anions, et à confirmer expérimentalement.

**VI. Dynamique de charge et décharge**

La caractéristique principale des supercondensateurs, par rapport aux batteries notamment, est leur grande puissance spécifique : ils se chargent ou se déchargent en quelques secondes. Dans la perspective d'optimiser ces dispositifs, il ne faut donc pas que l'augmentation de la capacité se fasse au détriment de la puissance. Les théories habituelles prédisent que les liquides sous confinement extrême sont fortement ralentis, ce qui disqualifie en principe l'utilisation de carbone micro- ou nanoporeux pour les supercondensateurs. Heureusement, l'effet n'est pas si dramatique que cela dans des structures à porosité interconnectée comme les carbones dérivés de carbures (CDCs), pour lesquels des temps caractéristiques de charge inférieurs à 20 secondes ont été observés pour les carbones présentant les pores de plus petite taille (0,8 nm en moyenne) [**4**]. Cette tendance est également confirmée par les mesures de spectroscopie d'impédance électrochimique, qui permet de sonder la résistivité de l'électrolyte à l'intérieur des pores dans le domaine des basses fréquences. Les valeurs obtenues (de l'ordre de 50 à 200 $\Omega$.cm pour des électrolytes organiques dans l'acétonitrile, à température ambiante) ne sont pas beaucoup plus élevées que celles mesurées dans les liquides (non confinés).

La charge rapide des supercondensateurs empêche l'utilisation des techniques *in situ* habituelles pour suivre leur évolution au cours du temps. Souvent, le temps nécessaire à l'enregistrement d'un spectre ou d'un diffractogramme est plus long que le temps de charge. Cependant, des techniques telles que la spectroscopie infrarouge ou le SAXS ont permis de suivre l'évolution de la structure à l'échelle des quelques secondes au cours de cycles charge/décharge pour un liquide ionique (1-éthyl-3-méthylimidazolium bis(trifluorométhylsulfonyl)imide) dans une électrode CDC ou encore une solution aqueuse de CsCl dans une électrode de carbone activé [22,23]. La RMN et l'imagerie de résonance magnétique (IRM) permettent également de suivre l'évolution des ions, à condition d'utiliser des cellules de mesure *in situ* avec un design particulier qui permet d'enregistrer la signature d'une seule électrode à la fois.

Enfin, la simulation moléculaire a permis de comprendre l'origine microscopique de la rapidité de charge. En particulier, l'étude des trajectoires de dynamique moléculaire permet d'extraire des propriétés de transport difficilement mesurables telles que les coefficients de diffusion des différentes espèces. Des travaux récents ont ainsi montré que les coefficients de diffusion des ions dans les électrolytes sont généralement diminués d'un ou deux ordres de grandeurs dans les électrodes (électrolytes confinés) par rapport au liquide (non confiné) [24], mais de fortes variations sont observées avec le remplissage des électrodes dans le cas des RTILs. Dans les CDCs, la connectivité du réseau de pore joue bien sûr un rôle important sur les propriétés de transport. A partir des résultats de simulation moléculaire, nous avons pu faire le lien avec un modèle de circuit électrique équivalent (voir **Figure 5**) et remonter ainsi au temps de charge pour une électrode réelle, de l'ordre de 1 à 10 s [25]. Ceci confirme les bonnes capacités prédictives de ces simulations, y compris du point de vue dynamique.

**VIII. Conclusion et perspectives**

La compréhension des mécanismes fondamentaux à l'échelle microscopique ces 5 dernières années fournit une base solide pour la conception de meilleurs supercondensateurs, en suggérant de nouvelles stratégies pour l'optimisation du stockage de la charge par une combinaison adéquate de structure des électrodes, d'ions et de solvant. L'utilisation à grande échelle des outils expérimentaux et de simulation développés pour établir ces mécanismes sera la clé du succès pour atteindre cet objectif.

Les développements méthodologiques ne sont bien sûr pas en reste, avec par exemple pour les simulations une nouvelle approche pour la prédiction des propriétés interfaciales en fonction de la différence de potentiel entre les électrodes, qui exploite les fluctuations d'équilibre de la charge des électrodes sous l'effet de l'agitation thermique dans l'électrolyte [26]. Nous avons ainsi pu faire le lien entre des pics de la capacité différentielle et des transitions, induites par la différence de potentiel, au sein de l'électrolyte adsorbé [27,28].

Du point de vue des applications, on pourra exploiter les possibilités offertes pour la fabrication d'électrodes de carbone dont on contrôle la porosité, la composition (par activation ou par dopage) ou la microstructure, ainsi que la large gamme de liquides ioniques et de solvants disponibles. Ces matériaux carbonés pourraient également jouer un rôle important dans d'autres contextes, par exemple avec des solutions aqueuses pour la récupération "d'énergie bleue", en exploitant la différence de salinité entre l'eau de mer et celle des rivières [29,30].

**Remerciements**

Les auteurs remercient leurs collègues Pierre-Louis Taberna et Barbara Daffos à Toulouse, Paul Madden à Oxford, ainsi que leurs anciennes doctorantes Céline Merlet, Clarisse Péan et Wan Yu Tsai. Ces travaux ont reçu le soutien de l'ANR (ANR-2010-BLAN-0933-02), de l'ERC (ERC grant agreement 102539), d'une chaire de la fondation Airbus Group et d'une chaire d'excellence de la Maison de la Simulation. Les résultats de simulation ont été obtenus grâce aux ressources de calcul du GENCI (projets c2013096728, x2012096728, x2014096728 et x2015096728), de PRACE (supercalculateur CURIE) et de l'EPSRC (supercalculateur HECTOR).

**Figures et légendes**

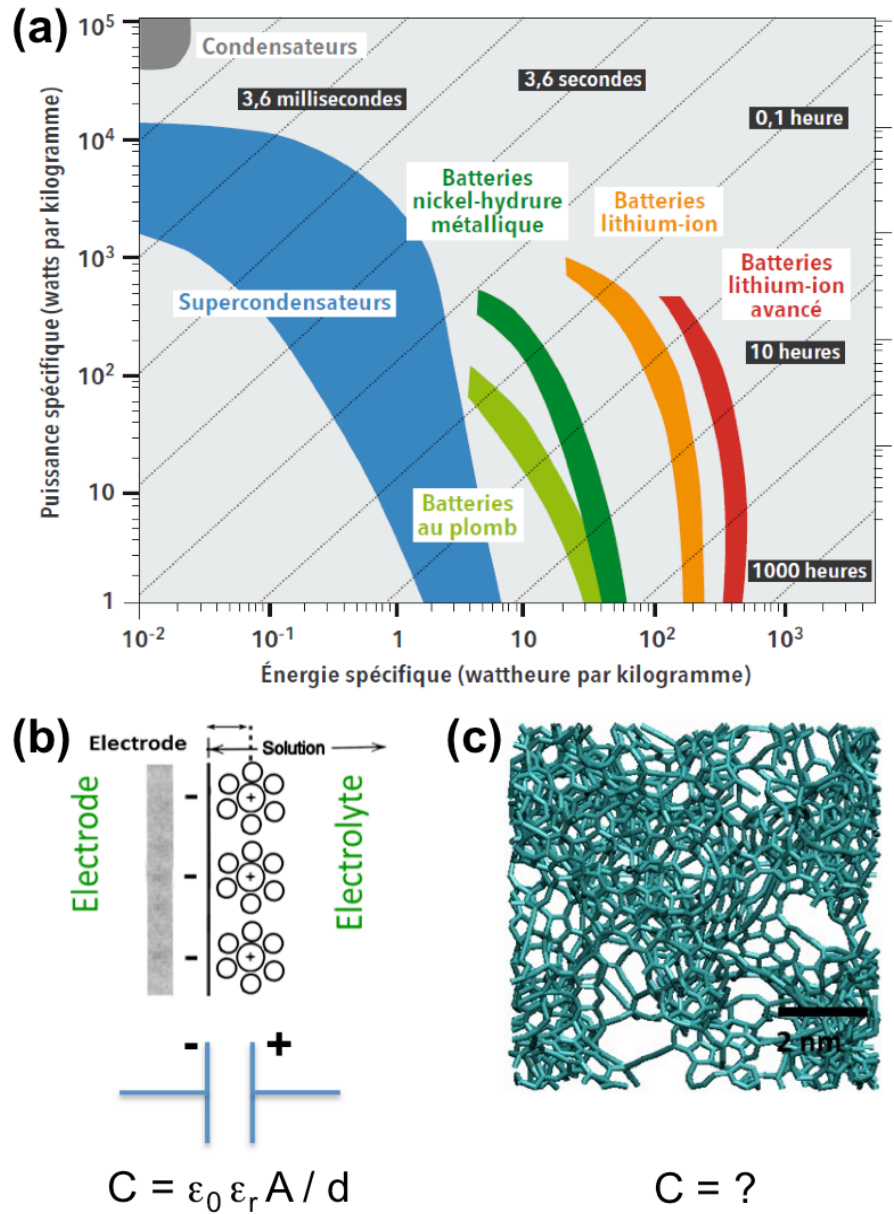

**Fig. 1.** (a) Le diagramme de Ragone représente les différents dispositifs de stockage de l'électricité en fonction de leur puissance spécifique (puissance par unité de masse) et de leur énergie spécifique (énergie par unité de masse). Les lignes diagonales indiquent le temps caractéristique de charge/décharge. (b) Représentation de la double couche électrochimique à la surface d'une électrode plane chargée négativement. Le modèle de condensateur plan prédit une capacité C proportionnelle à l'aire $A$ de l'électrode et inversement proportionnelle à la distance $d$ de séparation des charges. (c) Dans le cas d'un carbone poreux de grande surface spécifique (ici supérieure à 1000 $m^2.g^{-1}$), la prédiction de la capacité est plus délicate.

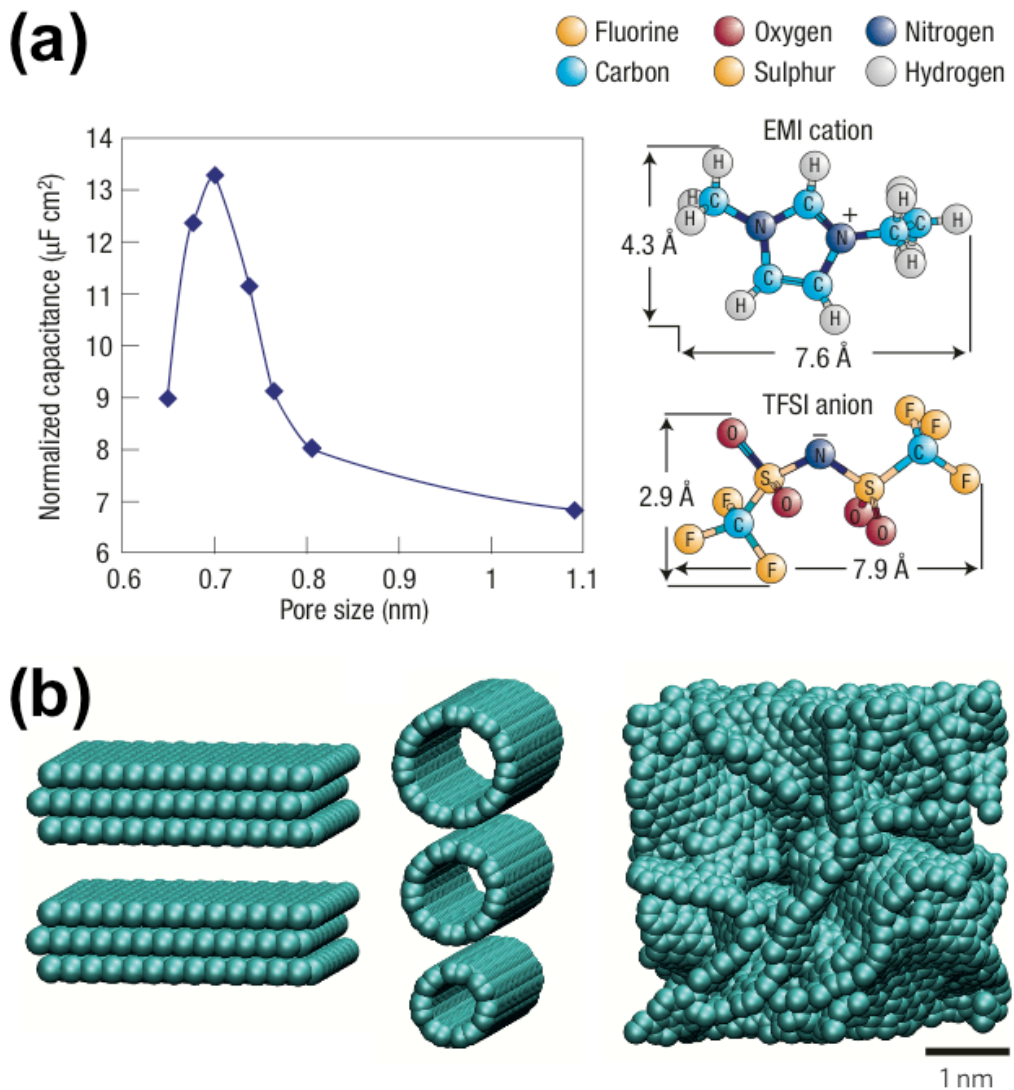

**Fig. 2.** (a) Capacité électrique par unité de surface obtenue pour le liquide ionique EMI-TFSI à 60°C avec des électrodes en carbones dérivés de carbures (CDC), en fonction de la taille moyenne des pores de l'électrode. Reproduit de Simon et Gogotsi, *Nat. Mater.* **2008**, *7*, 845-854 [5] avec la permission de Nature Publishing Group. (b) Exemples de structure de carbones: pores réguliers (fentes ou nanotubes) ou désordonnés (CDC). Reproduit de Salanne *et al.*, *Nature Energy*, **2016**, *1*, 16070 [6] avec la permission de Nature Publishing Group.

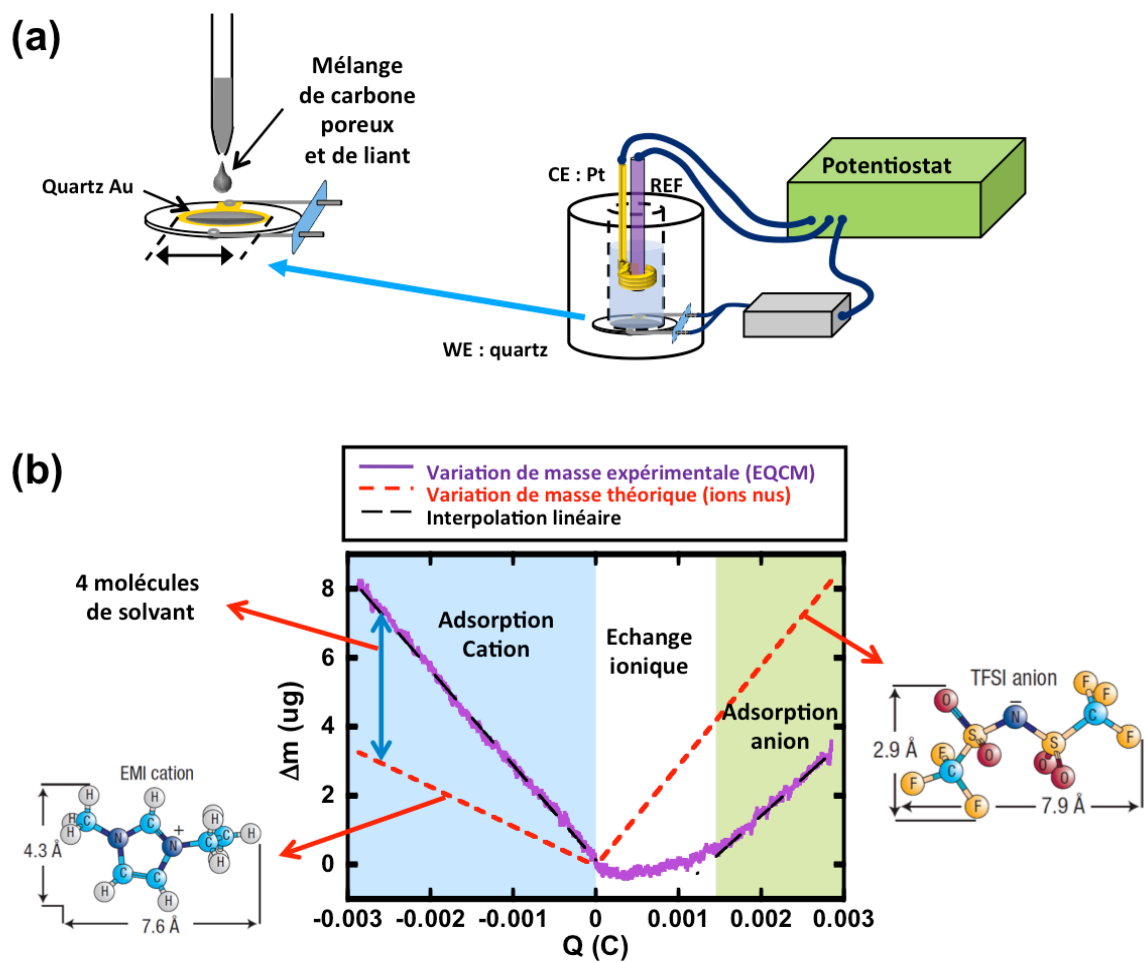

**Fig. 3.** (a) Schéma d'une microbalance à quartz électrochimique (EQCM). L'électrode de travail (WE) est déposée sur un quartz piézoélectrique qui permet de mesurer la variation de masse au cours de la charge (ici la contre-électrode CE est en platine). (b) La comparaison des résultats expérimentaux à la loi de Faraday (Equation 2) pour des ions nus ou solvatés permet de déduire les mécanismes qui interviennent en fonction de la polarisation de l'électrode.

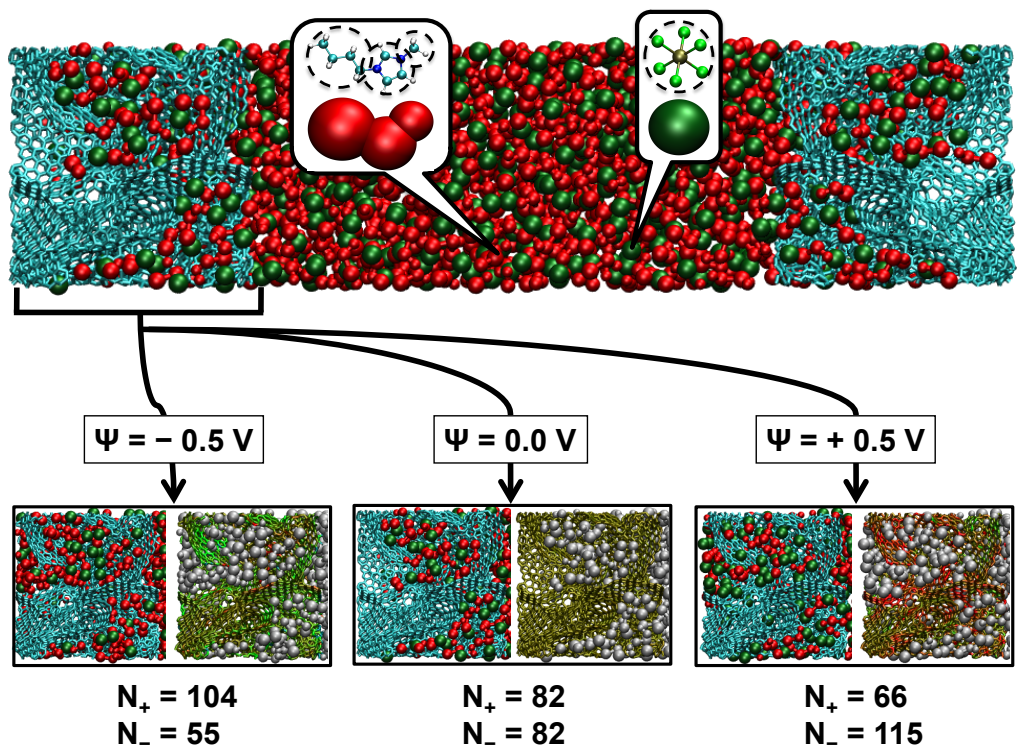

**Fig. 4.** Simulation moléculaire d'un supercondensateur constitué de deux électrodes nanoporeurses de Carbone Dérivé de Carbure (CDC), maintenues à une différence de potentiel constante, et d'un électrolyte liquide ionique à température ambiante. Les ions du liquide ionique, l'hexafluorophosphate de butyl-méthyl-imidazolium (BMI-PF$_6$), sont décrits par un modèle à "gros grains" (trois sites pour le cation, en rouge, un seul pour l'anion, en vert). Pour une différence de potentiel nulle ($\Psi$=0.0V), il y a autant de cations que d'anions dans chaque électrode, et la charge de ces dernières est nulle. Pour une différence potentiel de 1V, il y a un excès de cations dans l'électrode négative ($\Psi$=-0.5V) et un excès d'anions dans l'électrode positive ($\Psi$=+0.5V). Dans les deux cas, la charge nette du liquide dans l'électrode est compensée par la charge de cette dernière. La charge locale de l'électrode (négative en vert, positive en rouge) est illustrée, dans chaque cas, sur la partie droite de la figure correspondante. Ce mécanisme d'échange d'ions entre les électrodes diffère radicalement de ce qui se passe près d'une électrode plane de graphite. Reproduit de Merlet *et al. Nat. Mater.*, **2012**, *11*, 306 [**21**] avec la permission de Nature Publishing Group.

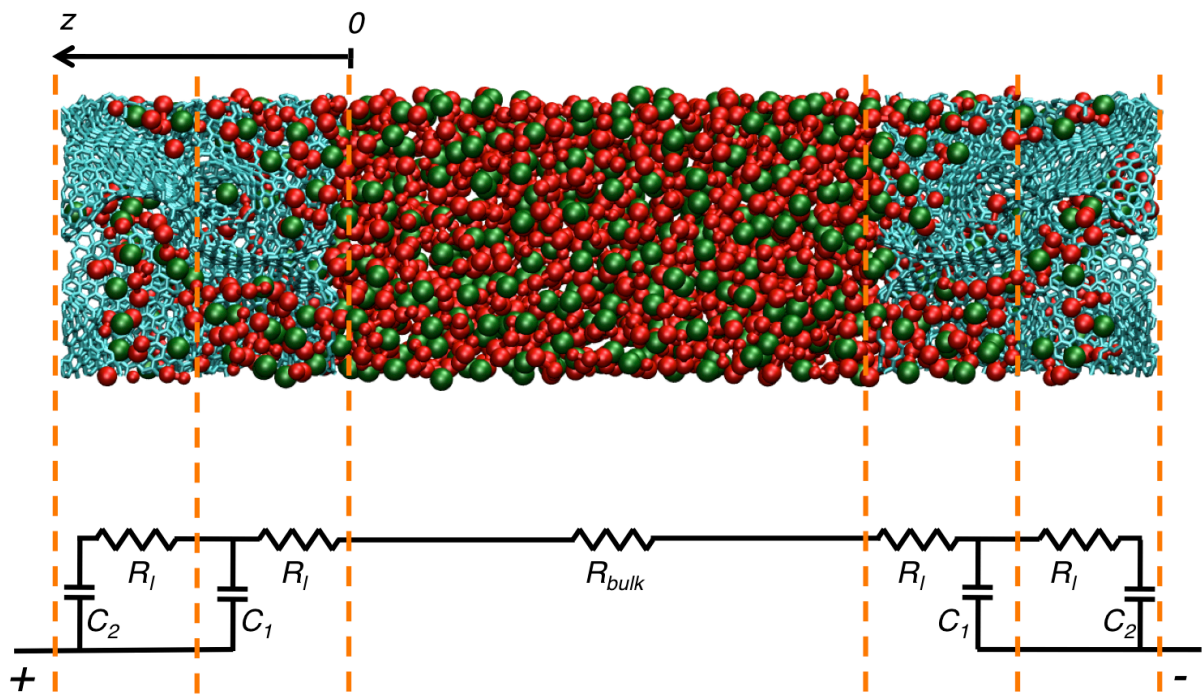

**Fig. 5.** La dynamique de charge peut être étudiée par simulation moléculaire en mesurant la charge des électrodes en fonction du temps lorsque l'on passe d'une différence de potentiel nulle à une valeur non-nulle (où l'inverse). Compte tenu de la taille nanométrique du système simulé, on ne peut directement comparer ces résultats aux données expérimentales. On peut par contre les analyser à l'aide d'un modèle de circuit électrique équivalent similaire à ceux utilisés par les expérimentateurs. Les paramètres correspondant à un modèle de ligne à transmission (résistance de l'électrolyte $R_{bulk}$, résistance et capacité par unité de longueur d'électrode $R_l$ et $C_i$) sont ainsi déterminés, ce qui permet d'extrapoler à un temps de charge pour un grain d'électrode de taille micrométrique (dans ce modèle, le temps de charge croît comme le carré de la taille) de quelques secondes, en accord avec la caractérisation électrochimique. Reproduit avec la permission de Péan *et al., ACS Nano*, **2014**, *8*, 1576 [25]. Copyright 2014 American Chemical Society.

**Encadré 1: Electrolytes pour les supercondensateurs**

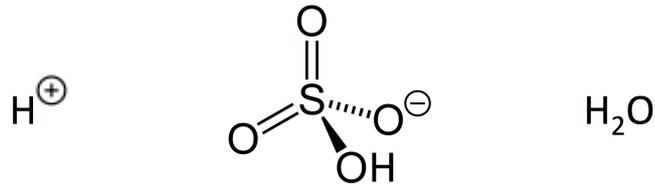
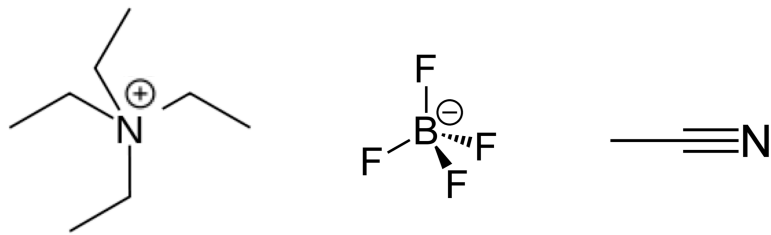
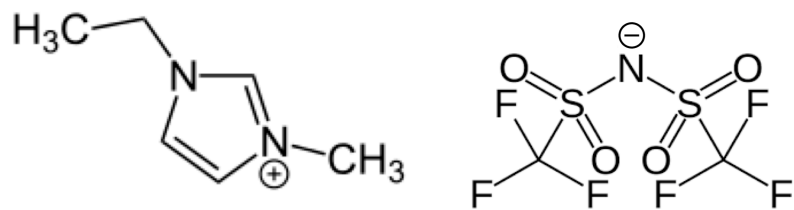

Trois types d'électrolytes liquides sont utilisés dans les surpercondensateurs actuels. Les électrolytes aqueux sont avantageux du point de vue environnemental et de la sécurité, mais ils possèdent une fenêtre électrochimique limitée. Cette dernière peut être fortement élargie en utilisant des solvants organiques avec des ions dissous, voire des liquides ioniques à température ambiante. Ces derniers possèdent cependant des conductivités ioniques relativement faibles.